\PassOptionsToPackage{dvipsnames}{xcolor}
\documentclass[conference]{IEEEtran}
\IEEEoverridecommandlockouts

\usepackage{cite}
\usepackage{amsmath,amssymb,amsfonts}
\usepackage{algorithmic}
\usepackage{graphicx}
\usepackage{textcomp}
\usepackage{xcolor}
\usepackage{balance}
\usepackage[all]{nowidow}
\usepackage{mathtools}
\usepackage{bm}
\usepackage{fnpct}
\usepackage{pifont}
\usepackage{booktabs}
\usepackage{makecell}
\usepackage[flushleft]{threeparttable}
\usepackage{multirow}
\usepackage{tabularx}
\usepackage{tikz}
\usetikzlibrary{patterns.meta}
\usepackage[hidelinks]{hyperref}
\usepackage[capitalise,noabbrev]{cleveref}
\usepackage{xspace}

\newcolumntype{Y}{>{\centering\arraybackslash}X}

\newcommand{\hatchbox}{%
\begin{tikzpicture}
    \draw[pattern={Lines[
        distance=0.8mm,
        angle=45,
        line width=0.15mm
       ]}, pattern color=black] (0,0) rectangle (0.3,0.3);
\end{tikzpicture}%
}

\newcommand{\emptybox}{%
\begin{tikzpicture}
    \draw (0,0) rectangle (0.3,0.3);
\end{tikzpicture}%
}

\newcommand{\ldotsbox}{%
\begin{tikzpicture}
    \draw (0,0) rectangle (0.3,0.3);
    \node at (0.05,0.05)[circle,fill,inner sep=0.25pt]{};
    \node at (0.05,0.117)[circle,fill,inner sep=0.25pt]{};
    \node at (0.05,0.183)[circle,fill,inner sep=0.25pt]{};
    \node at (0.05,0.25)[circle,fill,inner sep=0.25pt]{};

    \node at (0.117,0.05)[circle,fill,inner sep=0.25pt]{};
    \node at (0.117,0.117)[circle,fill,inner sep=0.25pt]{};
    \node at (0.117,0.183)[circle,fill,inner sep=0.25pt]{};
    \node at (0.117,0.25)[circle,fill,inner sep=0.25pt]{};

    \node at (0.183,0.05)[circle,fill,inner sep=0.25pt]{};
    \node at (0.183,0.117)[circle,fill,inner sep=0.25pt]{};
    \node at (0.183,0.183)[circle,fill,inner sep=0.25pt]{};
    \node at (0.183,0.25)[circle,fill,inner sep=0.25pt]{};

    \node at (0.25,0.05)[circle,fill,inner sep=0.25pt]{};
    \node at (0.25,0.117)[circle,fill,inner sep=0.25pt]{};
    \node at (0.25,0.183)[circle,fill,inner sep=0.25pt]{};
    \node at (0.25,0.25)[circle,fill,inner sep=0.25pt]{};
\end{tikzpicture}%
}

\newcommand{\sysnameplain}{duet instrumentation}
\newcommand{\Sysnameplain}{Duet instrumentation}
\newcommand{\sysname}{\sysnameplain\xspace}
\newcommand{\Sysname}{\Sysnameplain\xspace}
\newcommand{\sysnamebold}{\textbf{\sysnameplain}\xspace}

\begin{document}

\title{\Sysnameplain: An Agentic Approach to Improving Sensitivity in Cloud Service Benchmarking%
    \thanks{}%
}

\author{%
    \IEEEauthorblockN{Sebastian Koch, Nils Japke, David Bermbach}
    \IEEEauthorblockA{%
        \textit{Technische Universit\"at Berlin \& Einstein Center Digital Future}\\
        \textit{Scalable Software Systems Research Group}\\
        \{sk,\,nj,\,db\}@3s.tu-berlin.de%
    }%
}

\maketitle

\begin{abstract}
Continuous cloud service performance benchmarking is essential for detecting performance bugs early before deploying them to production.
However, detecting performance regressions using application benchmarks, which usually treat the system under test as a black box, is challenging due to variable I/O calls or changing performance characteristics of the underlying cloud infrastructure.
Microbenchmarks are often more sensitive and accurate, but also more time-consuming to implement and run.
Further, they do not capture the performance of the integrated system as a whole.
A comprehensive performance assessment therefore typically requires a combination of both approaches.

To address the shortcomings of application benchmarks, we propose \sysnamebold, a novel benchmarking paradigm enabled by the recent advancements of large language model (LLM) code understanding.
The idea is to analyze code changes between two consecutive application versions, and measure performance differences directly at performance-relevant changes during a synchronized benchmark of both application versions, uncovering performance changes with higher sensitivity.
We design a system that reliably automates the assessment and instrumentation of performance-relevant code changes between the two application versions.
In experiments with a realistic testbed application offering configurable performance regressions, we find that our prototype achieves 58\% precision, 93\% recall, and 71\% specificity (averaged across tasks) when comparing the generated instrumentation against the ideal instrumentation, with a line-distance threshold of five.
In the downstream application benchmark, we find that our prototype can detect performance regressions at up to 5$\times$ lower injected severity compared to a traditional duet application benchmark, while preserving similar A/A latency distributions (Hellinger distance 0.048--0.080).
\end{abstract}


\section{Introduction}
\label{sec:introduction}
With the increasing complexity and distributed nature of cloud-native software systems, it is challenging to catch performance regressions before they degrade user experience or violate service-level agreements (SLAs)~\cite{schermann2016bifrost}.
Especially in a fast-paced software development environment with frequent releases, developers should be alarmed early and continuously about emerging performance regressions to effectively mitigate performance-related issues.

Continuous cloud service benchmarking is a well-established practice for catching performance bugs early and for deriving knowledge about a system's performance characteristics under realistic stress scenarios~\cite{continuous_benchmarking_grambow,continuous_benchmarking_daly_2021,leitner2016patternschaosstudy,Bermbach_book_2017}, yet existing methods often fall short in simultaneously optimizing for efficiency, sensitivity, and relevance:
State-of-the-art cloud service benchmarking evaluates a system from two perspectives: a client-centric perspective using application benchmarks and a system-internal perspective using microbenchmarks.
Each offers unique insights but also comes with limitations and trade-offs.
Since application benchmarks take the perspective of a client, they are typically more business-relevant as they assess the end-to-end performance of an integrated system.
On the downside, they are insensitive to small performance changes and usually do not support root cause analysis.
Microbenchmarks, on the other hand, are more accurate and sensitive in detecting performance changes~\cite{Japke_2023} and allow developers to pinpoint root causes~\cite{Grambow_2023}.
As a downside, they are often incomplete or missing altogether from real software projects~\cite{missing_performance_tests} and are expensive to run~\cite{Grambow_2023,Japke_2023}.
Additionally, microbenchmarks fail to capture performance regressions in the context of the whole system, i.e., a seemingly minor performance change may compound to a massive regression if called frequently in the application.
Thus, both benchmarking types are traditionally required for a comprehensive assessment.

Both benchmarking approaches also suffer from performance variability in the underlying cloud infrastructure~\cite{leitner2016patternschaosstudy,Schirmer_2023,Laaber_2019}.
Benchmarking best practices, hence, call for techniques such as Duet Benchmarking~\cite{Bulej_2020_Duet_Benchmarking,Grambow_2023} for application benchmarks or RMIT~\cite{RMIT} coupled with bootstrapping~\cite{Davison_Hinkley_1997} for microbenchmarks~\cite{Laaber_2019,schirmer2024elastibench,Grambow_2023}.

To highlight the challenges of current cloud service benchmarking approaches, consider the following motivating example:
A cloud application provides an endpoint for retrieving the most popular movies in a region.
In a new commit, a developer adds work to an internal ranking routine (e.g., extra iterations in the algorithm).
Under cloud noise and highly variable I/O, this extra CPU work can be masked in end-to-end latency, thus, being invisible to the application benchmark.
At the same time, a microbenchmark may isolate the routine, but it cannot reproduce the end-to-end impact of concurrency, caching, and other factors that control end-to-end performance.

\begin{figure}[t]
    \centering
    \includegraphics[width=0.8\columnwidth]{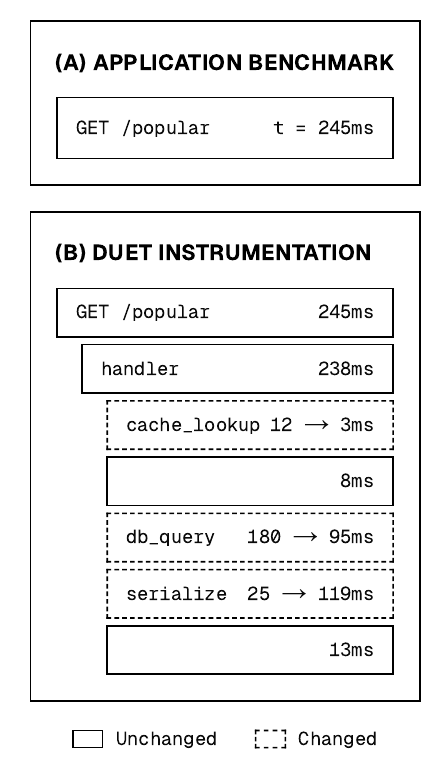}
    \caption{In comparison to a black-box application benchmark, \sysname adds change-localized measurements to identify sub-routine performance changes while keeping end-to-end relevance.}
    \label{fig:instrumentation-overview}
\end{figure}

In this paper, we propose \sysnamebold, a novel approach for closing the gap between both benchmark types which retrofits support for root cause analysis to client-centric application benchmarks.
For this, we automatically instrument the system under test (SUT), i.e., insert lightweight performance monitoring hooks.
Due to the performance impacts of tracing, we only do this for \emph{relevant} code segments, leveraging information on code changes from a version control system such as Git and large language models (LLMs) for interpretation.
To further reduce tracing impacts, we instrument \emph{both} versions and run them in a duet benchmarking setup, thus, offsetting instrumentation impacts similar to noise in XLR cables.
This way, \sysname allows us to measure the performance change resulting from relevant code segments directly at the change itself coupled with information on end-to-end performance changes; see \cref{fig:instrumentation-overview} for an overview and example.

We summarize our contributions as follows:
\begin{itemize}
    \item We propose the concept of \sysname, which integrates change-localized measurements into duet application benchmarks to improve sensitivity without requiring a dedicated microbenchmark suite. (\cref{sec:system-design})
    \item We design and implement an automated instrumentation pipeline that selects potentially performance-relevant code changes between two application versions and inserts lightweight measurement hooks to support downstream analysis. (\cref{sec:system-design})
    \item We use our prototype to study performance regressions in a realistic testbed application and show that \sysname can detect performance changes at up to 5$\times$ lower injected severity compared to classical duet benchmarking. (\cref{sec:evaluation})
\end{itemize}

\section{Background and Related Work}
We briefly summarize the background and related work to frame the problem and our approach.

\subsection{Background}
\label{sec:background}
First, we introduce background on cloud service benchmarking, performance change detection, and observability concepts.

\subsubsection{Cloud Service Benchmarking}
As already discussed in the introduction, cloud service benchmarks traditionally consist of two major types:

Similar to system and integration tests, \textbf{Application Benchmarks} take a client-centric perspective by deploying the application in a production-like, realistic environment and stressing the external interfaces, such as HTTP endpoints, with a workload that represents the expected production load.
The goal is to assess the integrated system as realistically as possible, as it is used by clients~\cite{Bermbach_book_2017}.

Similar to unit tests, \textbf{Microbenchmarks} take an internal perspective, repeatedly calling a single function or a small set of functions in isolation to measure the execution time for each run.
By assessing small subsets of the system, the objective is to find performance bugs early with higher sensitivity before they compound into noticeable regressions for the client~\cite{Laaber_2019,costa2019s}.
In practice, microbenchmarks usually come in suites of possibly hundreds of microbenchmarks, which can take days to execute~\cite{schirmer2024elastibench,Grambow_2023}.
Microbenchmarks should also be as realistic as possible, but as they only provide a lens on a small part of the system, their workloads inherently lack the end-to-end insights of application benchmarks.

Benchmarking cloud services is a challenging task since the cloud can provide compute instances with highly variable performance characteristics for different experiment runs~\cite{Bermbach_book_2017,Laaber_2019}.
Effects such as noisy neighbors that impact the CPU steal time or context switches can alter the performance of the system under test, making it difficult to attribute performance changes to actual code differences.
Following good benchmarking practices, a benchmark should produce predictable and repeatable results, therefore new approaches have evolved that at least partially mitigate issues associated with the variability of the cloud.

Traditionally, benchmarks run sequentially while \textbf{duet benchmarking}~\cite{Bulej_2020_Duet_Benchmarking} executes two service versions in parallel on the same machine with isolated resources.
The idea is that external performance noise affects both service versions similarly due to the parallel execution in the same environment, and therefore cancels out when calculating the relative performance differences.
Previous research indicates that duet benchmarking provides more accurate results~\cite{Bulej_2020_Duet_Benchmarking,Grambow_2023}.
To achieve the best possible overlap of the influencing factors between the two versions, the execution needs to be synchronized.
This means synchronizing the start time of each request for both versions and setting the same resource limits.

\subsubsection{Performance Change Detection}
\begin{figure}[t]
    \centering
    \includegraphics[width=\columnwidth]{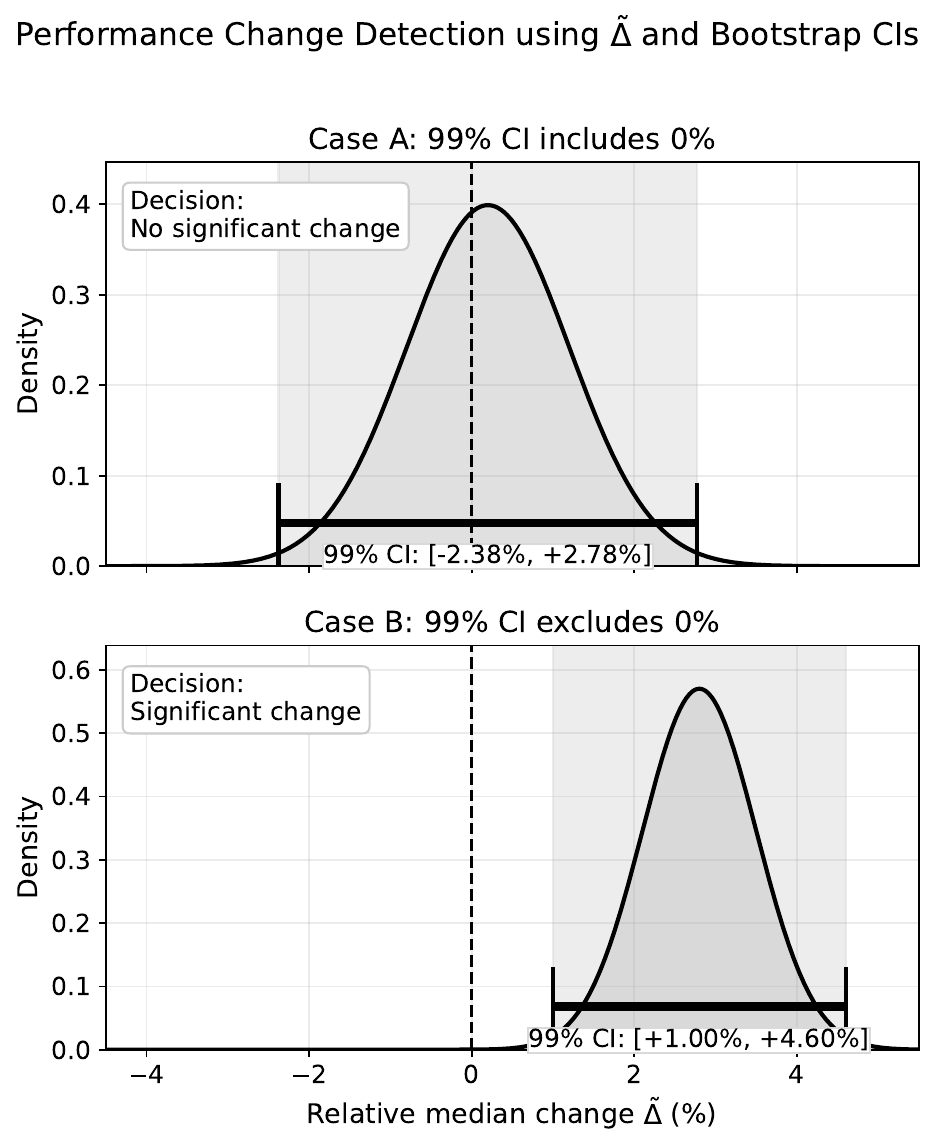}
    \caption{Performance change detection with bootstrap confidence intervals. If the confidence interval does not include $0\%$, we report a significant change.}
    \label{fig:perf-change-detection-demo}
\end{figure}

For this paper, we focus on detecting performance changes between two versions of a service.
We are particularly interested in the sensitivity of the detection, i.e., at what severity level we can detect a performance change.
While there are different methods for determining whether a service experienced a significant performance difference, we use the relative median of performance measurements in combination with bootstrapping confidence intervals as inspired by previous research~\cite{Japke_2023,japke2025muoptimestaticallyreducingexecution,Grambow_2023}.
For paired measurements $(x_i^{(A)}, x_i^{(B)})$, we compute the relative change
$\Delta_i = \left(\frac{x_i^{(B)}}{x_i^{(A)}} - 1\right)\cdot 100\%$ and define the relative median as
$\tilde{\Delta} = \mathrm{median}(\Delta_i)$.
A \emph{paired measurement} is a synchronized run of version $A$ and version $B$.
Bootstrapping confidence intervals~\cite{Davison_Hinkley_1997} is a non-parametric method that does not assume a specific distribution of the underlying data.
The method is based on resampling the data with replacement to approximate the sampling distribution.
We compute the confidence interval using bootstrapping for $\tilde{\Delta}$ and assess whether it intersects the value $0$.
As illustrated in \cref{fig:perf-change-detection-demo}, if the interval does not include $0$, we conclude that the performance between the two versions is significantly different.

\subsubsection{Observability \& Instrumentation}
We define instrumentation as the process of adding measurement code to the system under test to monitor its internal state, while observability is the ability to observe that state from measurable outputs such as logs, metrics, and traces.

In cloud service benchmarking, measurements are often heterogeneous, i.e., the data model of measurement results may differ between benchmark designs.
To have a unified result format, we adapt the OpenTelemetry\footnote{\url{https://opentelemetry.io/}} model for our benchmarking experiments.
OpenTelemetry evolved as a standard in the cloud-native space for observability and instrumentation and we, therefore, borrow the concepts of traces and spans for our performance measurements.

In OpenTelemetry, traces are a collection of spans that represent a single request or transaction in a cloud system.
Spans are the individual units of work that make up a trace.
A span has a name, usually corresponding to the operation being performed, a start and end time of the operation, and additional attributes associated with the operation.
Further, a span can be a child of another span, allowing the construction of an operation tree in a trace.

\subsection{Related Work}
\label{sec:related-work}
In our work, we build upon the findings of Japke~et~al.~\cite{Japke_2023} that microbenchmarks are more accurate and sensitive in detecting performance changes than application benchmarks.
We leverage this insight and instrument application benchmarks in a way that resembles the more sensitive measurement nature of microbenchmarks, enabling better root-cause analysis.
At the same time, we retain the unique advantages of application benchmarks.

To increase accuracy, we employ duet benchmarking, which empirically improves benchmarking accuracy~\cite{bulej2019initial,Bulej_2020_Duet_Benchmarking,japke2025isolation}.
In our setup, duet benchmarks also mitigate the impact of measurement overhead.

Previous work recognizes the need to detect code-level performance regressions for every commit~\cite{Reichelt_2022_Automated_Code_Level_Performance_Changes,Chen_JIT_Performance_Regression}, but previous methods rely on transforming unit tests into microbenchmarks.
This transformation limits relevance since unit tests often use mocks and rarely cover the entire code base.
Determining regressions on the integrated system is not feasible with such an approach either.
To enable benchmarking on ideally every commit during software development, there is ongoing research regarding benchmarks as part of CI/CD pipelines~\cite{waller_including_2015,javed_perfci_2020,daly_industry_2019,continuous_benchmarking_grambow,LaaberMBEval,japke2025cicd}.

BeFaaS~\cite{grambow2021befaasapplicationcentricbenchmarkingframework} is a benchmarking framework that uses automated code instrumentation to support drill-down analysis for application benchmarks.
While sharing similar objectives, our approach differs fundamentally.
We design \sysname around duet benchmarking and, therefore, need to measure symmetrically between two service versions.
This requires a more sophisticated approach, which we achieve by leveraging the reasoning capabilities of recent LLMs.
Our instrumentation is context-aware and aims to measure a minimal but meaningful subset of code changes, which scales well for complex applications.
In BeFaaS, the instrumentation aims to cover the entire code base without context understanding, which does not scale well.

Chen~et~al.~\cite{Chen_JIT_Performance_Regression} present a method to identify whether a commit might exhibit performance regressions.
We could use their method in our code change evaluation in the future, as a potentially more scalable alternative to LLM-based assessments or as a preselection step.
Nevertheless, due to the complexity of code changes, more powerful methods might be better suited to assess performance relevance.

A possible alternative to our approach is live testing as proposed in Bifrost~\cite{schermann2016bifrost}, which gradually rolls out a new version in production while monitoring metrics in regard to performance or error rates.
Live testing has inherent advantages, especially in combination with OpenTelemetry-based instrumentation, as it uses real production load.
However, the instrumentation is not change-localized and therefore less sensitive to performance changes.
The root-cause localization may be harder to achieve as well.
Further, \sysname runs in the CI/CD pipeline before promoting changes to production and provides developers with earlier feedback.

\begin{figure*}[t]
    \centering
    \includegraphics[width=0.8\textwidth]{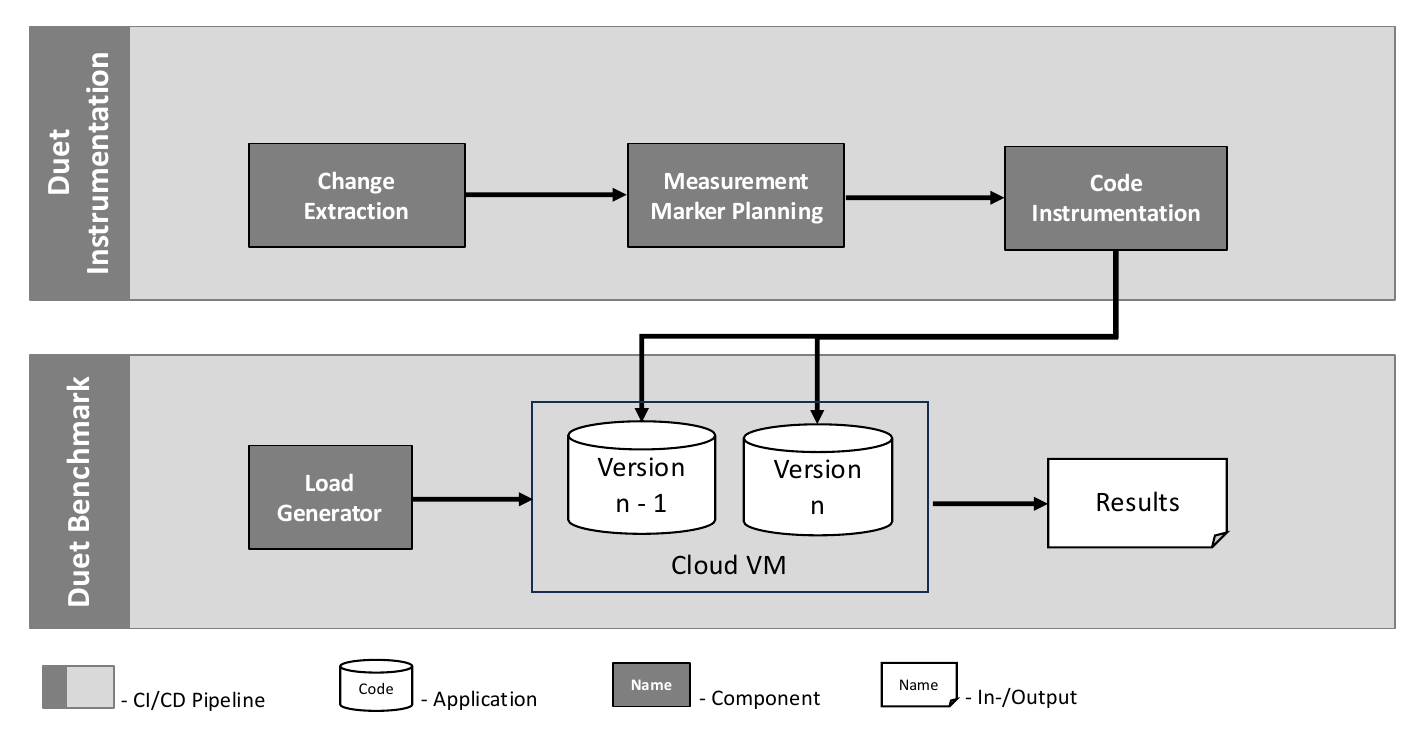}
    \caption{Architecture of \sysname. We add change-localized performance measurements to the two application versions and prepare them for a duet benchmark. Subsequently, we run the applications in a duet benchmark and analyze the results.}
    \label{fig:workflow}
\end{figure*}

\section{System Design: Automated Duet Instrumentation}
\label{sec:system-design}
We present \sysname, a novel approach for cloud service benchmarking that automatically adds change-localized performance measurements to two application versions and subsequently runs a duet benchmark to evaluate relative performance differences.
Given a diff between versions $v$ and $v'$, we identify code changes that likely affect performance and measure the affected code regions in both versions.
Running the instrumented versions in parallel with a synchronized workload generates traces that correlate performance regressions with specific code changes and support root-cause analysis.

As shown in \cref{fig:workflow}, \sysname works in four steps:
(1) \emph{Change Extraction} identifies the changed files and their corresponding diffs from different change sources, such as Git, creating a \emph{change set} for the next stage.
(2) \emph{Measurement Marker Planning} employs an LLM to assess the changes and proposes likely performance-relevant changes with affected contexts, referred to as \emph{marker plans}.
(3) \emph{Code Instrumentation} converts the marker plans into measurements for both versions.
(4) The \emph{Duet Benchmarking Procedure} runs both instrumented versions in parallel with isolated resources and sends a synchronized request workload.

We design \sysname to run within CI/CD to continuously inform developers about emerging performance bugs in evolving codebases, i.e., the pipeline must run without human intervention.

\subsection{Change Extraction}
In the first step, we receive two application versions as input, e.g., from a Git diff.
We extract the changed files and their diffs, forming a change set, which is the input for the next step.
The change set provides an abstraction of the change source to enhance the system's modularity.
Beyond version control systems, we can support other sources, such as JSON file diffs, as long as we provide adapters for these.
Since Git diffs are a common change source, we transform inputs into a Git-like format for the downstream LLM to process, because we expect LLMs to work particularly well with this format.

\subsection{Measurement Marker Planning}
The main challenge of \sysname is to find the minimum number of measurement points that locate all relevant performance regressions while keeping the instrumentation overhead low.
Optimizing for both high recall and precision represents a trade-off.
Therefore, we control the sensitivity to performance changes with a parameter that maps to textual instructions on what to consider performance-relevant.
Those textual instructions can vary for different applications and the specific implementation of \sysname.
A more sensitive system will achieve higher recall, but may exhibit lower precision.

Consequently, we seek Pareto-efficient solutions for a given sensitivity.
We address this problem by employing an LLM agent to assess changes and generate marker plans for performance-relevant changes, i.e., a list of spans with their start/end, names, and attributes such as operation categories or other meta-information.
We provide the mapped instructions of the sensitivity parameter to the agent as an input in the prompt.

The agent itself follows the ReAct paradigm~\cite{Yao_2023_ReAct} and turns reasoning decisions into actions taken as code.
The final output is a JSON object that we parse against the predefined schema for a marker plan.
The schema includes the span start and end, along with optional context attributes as mentioned before, which are useful for root-cause analysis.
In case the agent fails to adhere to the schema, we skip the change.
Importantly, we instruct the agent to consider exit points such as return statements or exceptions to end a span.

Because the agent's output is marker plans following a strict schema, we can verify the plans and deterministically convert them into spans, preventing the LLM from directly editing the application code.

To reduce instrumentation latency and costs, we cache marker plans using the change fingerprint as the cache key.

Please note that -- due to the fast pace of LLM development -- we consider specific prompts and LLM versions to be part of the prototype and not the overall \sysname approach.
More details in this regard can hence be found in \Cref{sec:evaluation}.

\subsection{Code Instrumentation}
For code instrumentation, we follow the widely adopted OpenTelemetry tracing model.
By following this model, we can use standard tracing tools for our measurements.

Because the OpenTelemetry SDKs can introduce non-negligible overhead~\cite{instrumentation_overhead}, we adhere to the OpenTelemetry data model and semantic conventions\footnote{\url{https://opentelemetry.io/docs/concepts/semantic-conventions/}} but design our own more lightweight SDK optimized for \sysname.
The objective is to minimize overhead while preserving compatibility at the level of trace/span concepts, identifiers, and common attributes.
Since we can miss performance-relevant changes with the automated instrumentation, we want to minimize interference with end-to-end latency through measurements.
Therefore, we omit most other OpenTelemetry features to achieve minimal overhead.

We insert measurements at the line ranges specified in the marker plans.
To support duet benchmarking, we write the instrumented versions of $v$ and $v'$ to disk.
Alongside the instrumented code, we store the run's metadata (e.g., the marker plan) to enhance traceability and reproducibility.

Because we want to directly compare both instrumented versions at the span level in an offline analysis step, we apply the plans symmetrically in both versions using the same names.
If we cannot ensure the symmetric insertion of a marker, we record the mismatch and skip it.
This preserves the duet principle, i.e., symmetric spans where possible, while ensuring robustness to additions, deletions, and refactoring.
Otherwise, we might skew the end-to-end latency of one version due to a larger number of measurements with the associated overhead.

After inserting a span, we syntactically check the applications.
If we cannot confirm a successful instrumentation, we skip the change and report the failure.
Through these design decisions, we ensure that we do not introduce any functional or breaking changes.

\subsection{Duet Benchmarking Procedure}
After instrumenting both versions, we deploy them to the same cloud VM with isolated resources, i.e., each application runs in its own container with dedicated CPU and memory.
We pin each container to unique physical CPU cores, with one core reserved for operating system tasks to prevent interference with application performance.
To synchronize the benchmark, we send the same requests to both applications simultaneously and measure the performance of each request, both internally with our instrumentation and externally on the client, as the overall request duration.

\section{Evaluation}
\label{sec:evaluation}
We evaluate our approach both through a proof-of-concept implementation following the system design from \cref{sec:system-design} and through a number of experiments with a testbed application.
In that regard, we use a realistic, purpose-built testbed application with performance regressions of adjustable severity across versions.
Our experiments consist of instrumentation, benchmarking, and analysis, in which we evaluate the instrumentation quality, performance change detection, and system overhead.
As our \emph{baseline}, we use the same testbed application and workloads but without change-localized instrumentation.
We make the prototype, testbed application, and experiment results publicly available.\footnote{The code and data are available at \url{https://github.com/s3bastiankoch/duet-instrumentation}.}

\subsection{Prototype Implementation}
We implement a Python-based command-line interface (CLI) tool that supports \sysname for TypeScript applications.
We chose TypeScript due to its popularity in the industry, being the most commonly used language on GitHub~\cite{octoverse2025}.
Nevertheless, we can extend our CLI to other languages.
The CLI takes the path to the application entry point as an input and instruments the code as described in \cref{sec:system-design}.

In the change extraction step, we support both Git and JSON input formats.
We use the JSON format for our experiments, as it resembles Git changes and also includes meta-information, such as ideal span boundaries for each change, which makes evaluation easier.

Our lightweight OpenTelemetry-inspired instrumentation code writes spans in batches to the local file system in an SQLite database to minimize interference with the benchmark.

For generating marker plans, we rely on a commercial LLM, namely GPT-5.2 by OpenAI.
We implement the agent using the smolagents library\footnote{\url{https://github.com/huggingface/smolagents}} and release the prompts and outputs in our open-source repository.

\subsection{Testbed Application}
As a testbed, we develop a service for movie search, recommendation, and ranking with workloads that stress the network, memory, and CPU, serving as a realistic application.
We use a purpose-built testbed rather than mining open-source systems for performance bugs because our sensitivity analysis needs adjustable ground-truth performance changes with known locations, so we can assess localization quality and regressions under controlled severity, i.e., we inject the same performance change across a 1\%--100\% severity range.
Mined performance bugs typically do not allow severity adjustment.
We design the service to reflect common cloud-service characteristics (REST API, database access, CPU-bound ranking/recommendation, and embedding-based search).
For movie recommendations, we adopt a collaborative filtering algorithm based on matrix factorization, historically employed by Netflix~\cite{hastie2014matrixcompletionlowranksvd}.
In the movie ranking, we implement a simplified version of the PageRank algorithm~\cite{pagerank}, as popularized by Google for ranking web pages.
Finally, our movie search uses a hybrid search, combining the established BM25 algorithm with a k-nearest neighbor (kNN) search over item embeddings generated by the \texttt{all-MiniLM-L6-v2} language model\footnote{\url{https://huggingface.co/sentence-transformers/all-MiniLM-L6-v2}}.
We implement the application in TypeScript with a RESTful API and a MongoDB database for storing movies, users, and preferences.
We prefill the MongoDB database with multiple movies, users, and preferences upon application start.
The RESTful API exposes three relevant endpoints:
one endpoint for searching movies with a query parameter, one for ranking movies for a specific region, and one for user-specific movie recommendations.
Each algorithm has a parameter that increases computational complexity, which we use to introduce repeatable performance regressions.

\subsection{Experiment Setup}
\label{sec:experiment-setup}
In our experiments, we evaluate three \emph{use cases} (U1--U3) and nine \emph{experiments} (E1--E9).

\paragraph{Use cases (U1--U3).}
We benchmark the three API endpoints of our testbed application, each stressing different resources:
(U1) movie search, (U2) movie ranking, and (U3) movie recommendation.

\paragraph{Injected regressions and severity levels.}
For each use case, we introduce a performance regression with adjustable severity as described in \cref{tab:use-cases}.
Different severity levels add additional effort to computational tasks.
We use the levels 1\%, 2\%, 3\%, 5\%, 10\%, 20\%, 50\%, and 100\% more effort, respectively.
To more closely resemble real-world code changes, the performance-relevant changes affect work in multiple downstream parts of the request.
We additionally inject non-performance-relevant comments and debug logs as noise.

\paragraph{Experiments (E1--E9).}
We run the following experiments:
\begin{itemize}
  \item \textbf{E1--E3}: 5 experiments per severity, where we only run the instrumentation step and quantify metrics in regard to the localization quality.
  \item \textbf{E4--E6}: Performance change detection benchmarks (3 runs per severity) comparing end-to-end latency against internal measurements between versions.
  \item \textbf{E7--E9}: Instrumentation overhead via A/A benchmarks with 0\% performance difference (5 runs), evaluating latency distributions between an instrumented and an uninstrumented version.
\end{itemize}
Within each block, experiments correspond to use cases in order: (U1,U2,U3).

For E1--E3, we repeatedly run the instrumentation to capture run-to-run variance of instrumentation quality under identical inputs.
For E4--E6, we follow the performance-change detection method from \cref{sec:background}.

We run all our experiments as duet benchmarks at similar times of the day in Google Cloud on \texttt{c4-highcpu-8} machines, with 8 vCPUs and 16 GB RAM, for the SUTs and \texttt{e2-standard-2} machines, with 2 vCPUs and 8 GB RAM, for the load generator in the \texttt{europe-west3-a} zone.
We separate the load generator from the SUTs to avoid interference with the application performance.

\begin{table*}[t]
  
  \centering
  \renewcommand{\arraystretch}{1.5}
  \caption{Use cases (U1--U3): endpoints and injected performance-relevant change.}
  \label{tab:use-cases}
  \begin{tabular}{@{}p{2cm} p{1cm} p{3cm} p{10cm}@{}}
  \toprule
  № Use case & № Span & Endpoint & Nature of Performance Change \\
  \midrule
  U1 & S1 & /search & Change to the limit parameter of the search, increasing the number of movies to consider during downstream operations. \\
  U2 & S2 & /top/region/:region/count & Random shuffle of internal list for random walk algorithm. \\
  U3 & S3 & /top/user/:userId/count & Regression in internal loop of matrix factorization with more iterations than necessary. \\
  \bottomrule
  \end{tabular}
  \end{table*}

\subsection{Results}
We report results for (1) instrumentation quality (E1--E3), (2) performance-change detection sensitivity (E4--E6), and (3) instrumentation overhead (E7--E9).

\subsubsection*{Instrumentation Quality}
We evaluate the instrumentation quality in multiple dimensions:
(1) How many of the relevant changes are we able to instrument? If we instrument them, how closely does the instrumentation align with the change context? (2) How many of the irrelevant changes are we able to skip?

We define a positive hit if a generated span is within $k=5$ lines of a performance-relevant affected context; this threshold accounts for non-executable lines (comments, blank lines, whitespace) that inflate the line distance but do not change the semantic placement.
Using this threshold, we report precision@5 (fraction of generated spans that are a positive hit), recall@5 (fraction of performance-relevant changes that have a positive hit), and specificity@5 (fraction of performance-neutral changes that have no positive hit).
We calculate the distributions across use cases to highlight instrumentation quality.

\begin{figure}[t]
    \centering
    \includegraphics[width=\columnwidth]{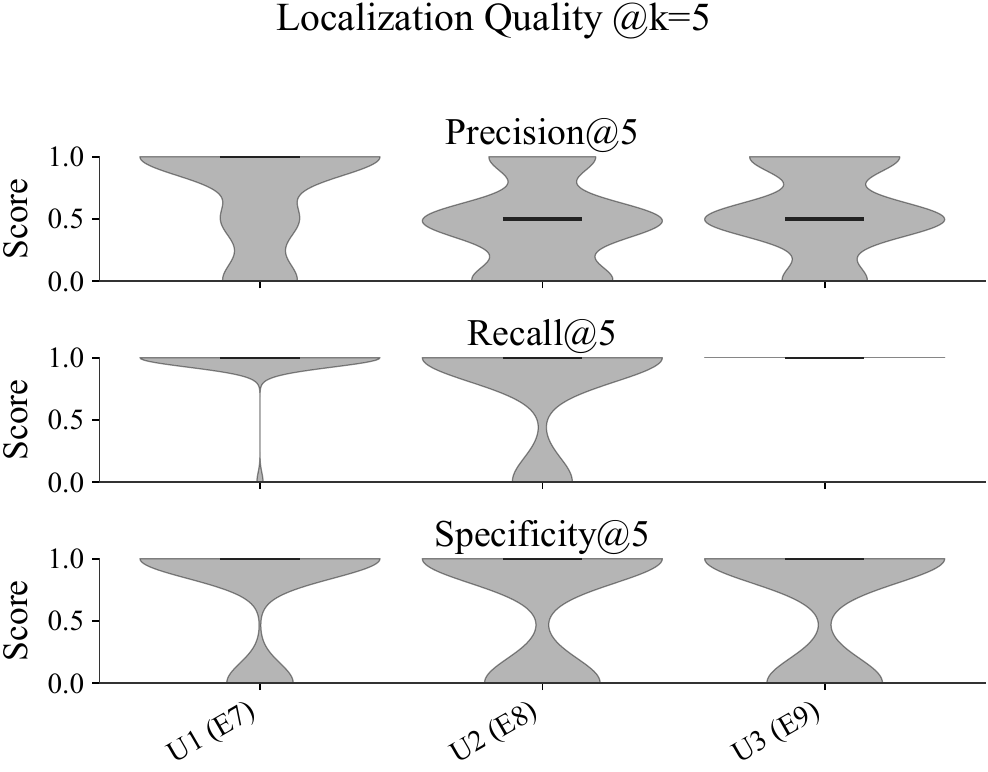}
    \vspace{-0.4em}
    \caption{Precision@5, recall@5 and specificity@5 distributions across use cases (U1--U3 / E1--E3). Median shown as horizontal line.}
    \label{fig:precision-recall-violin}
    \vspace{-0.8em}
\end{figure}

\cref{tab:localization-quality} depicts the results of our experiments for the average localization quality at $k=5$ lines, while \cref{fig:precision-recall-violin} shows the distributions.

\begin{table}[t]
    \centering
    \caption{Average localization quality at $k=5$ lines.}
    \label{tab:localization-quality}
    \begin{tabular}{@{}lccc@{}}
    \toprule
    Use Case & precision@5 & recall@5 & specificity@5 \\
    \midrule
    U1 & 0.71 & 0.98 & 0.78 \\
    U2 & 0.46 & 0.80 & 0.68 \\
    U3 & 0.56 & 1.0 & 0.68 \\
    \bottomrule
    \end{tabular}
\end{table}

Traditionally, recall and precision represent a trade-off, i.e., increasing one decreases the other.
Since our objective is to detect performance changes early and increase sensitivity, we aim for a high recall, willingly accepting false positives.
The results indicate that we detect performance-relevant changes with high recall, but that we can still improve the precision, as the instrumentation introduces additional measurement points per change, which leads to false positives.
As we later explore in \cref{sec:overhead}, those false positives do not impact the external interfaces much and are therefore tolerable.
Generally, we observe that our prototype is overly sensitive in adding measurement points per change.
This behavior is also reflected in the specificity metric, where we add measurements to performance-neutral changes.

\begin{figure}[t]
    \centering
    \includegraphics[width=\columnwidth]{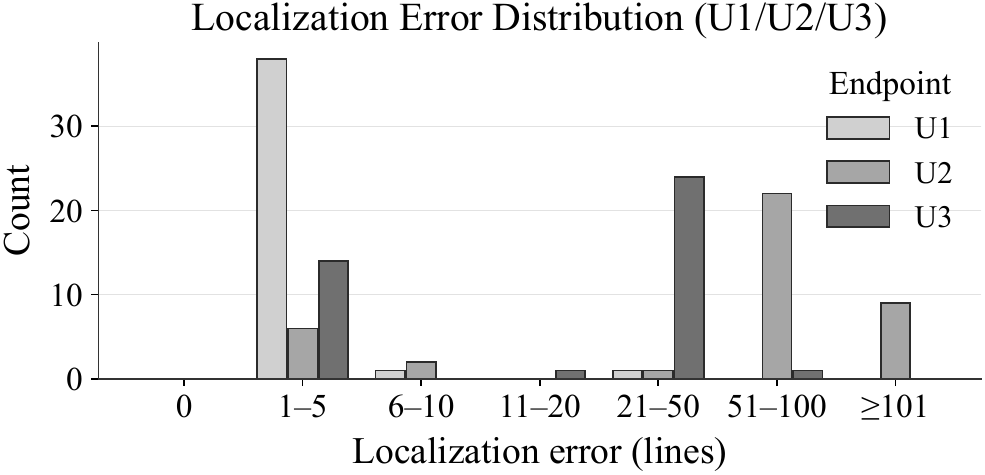}
    \caption{Average localization error in lines per use case.}
    \label{fig:localization-error}
\end{figure}

Beyond precision, recall, and specificity, we additionally evaluate the localization quality in terms of the distance between the planned perfect span and the actual spanned context, i.e., the localization error.
We define the perfect span as a human-provided span that most closely matches the affected context of a performance-relevant change.
We report the average localization error for each endpoint/use case as depicted in \cref{fig:localization-error}.

Across U1--U3, we find that the average localization error is often within the range of 1--5 lines, indicating accurate localization.
While U1 is mostly very accurate, U2 and U3 show a gap in the range of 6--20 lines and then exhibit higher localization errors.
This is driven by our prototype's tendency to add multiple measurement points per file to gather more information about a broader context, which we define as unnecessary overhead compared to the human-provided span.
As described in \cref{sec:discussion}, choosing the right balance of sensitivity and overhead is challenging and requires further research.

Notably, we choose $k=5$ lines as a relatively strict threshold for the localization quality, but our prototype is typically able to detect performance changes with higher sensitivity compared to the traditional approach even if the instrumentation is not within this threshold.

Since the LLM agent cannot directly edit code, and we enforce syntax checks in the pipeline, we observe no syntax errors in the instrumentation throughout our experiments.
All instrumentation plans are valid.

\subsubsection*{Performance Change Detection}
We compare the results of our prototype with the results of a traditional duet benchmark.
We evaluate the sensitivity of our system by measuring if we can report a performance change for a given severity level.
For scenarios where the instrumentation introduced multiple measurement points per file, we report a performance change if at least one of the measurement points shows a significant change following the method described in \cref{sec:background}.
Generally, we expect internal measurements to be more sensitive to performance changes than external measurements.
In these experiments, we validate that the measurements are not only accurately placed as previously described, but also sensitive to performance changes.

\cref{tab:results} depicts the results of our experiments with $3.79$M measurements gathered across all experiments, including span measurements; endpoint-only (client-side) measurements account for $197$k.

For U1 (E4), we detect performance changes at the $10\%$ severity level, with $CI = [5.09\%, 24.50\%]$, while the traditional approach only detects changes at the $50\%$ severity level, with $CI = [1.48\%, 12.32\%]$.
For U2 (E5), we detect performance changes at the $2\%$ severity level, with $CI = [15.50\%, 19.47\%]$, while the traditional approach detects changes similarly at the $2\%$ severity level, with $CI = [16.03\%, 25.54\%]$.
However, the traditional approach for U2 (E5) is quite volatile, reporting performance improvements at higher severity levels, which undermines its reliability.
For U3 (E6), we detect performance changes already at the $1\%$ severity level, with $CI = [5.78\%, 6.22\%]$, while the traditional approach only detects changes at the $5\%$ severity level, with $CI = [2.70\%, 14.06\%]$.
U2 (E5) is the only use case where we see similar detection capabilities for both approaches, due to the nature of the regression.
The regression strongly impacts the overall request duration and is therefore easier to detect on the external interface.

The results suggest that our approach can detect performance changes sooner than the traditional approach.
We detect pronounced changes in the overall request duration similarly well in both approaches.
Smaller changes, however, are not easy to detect on the external interface, but are equally important to find since they can lead to a performance drift over time.
Even if the traditional approach finds performance changes at the same severity level, the confidence intervals are typically wider, indicating a higher uncertainty in the results.

\begin{table}[t]
    \centering

    \caption{Results for the performance change detection experiments.}
    \begin{threeparttable}
    \begin{tabularx}{\columnwidth}{>{\centering}p{0.6cm}YYYYYYYp{0pt}YYYY}
    \toprule
    & \multicolumn{3}{c}{\makecell{Instrumentation\\Spans}} & & \multicolumn{3}{c}{\makecell{Application\\Endpoints}} \\\cmidrule(l{1pt}r{1pt}){2-4} \cmidrule(l{1pt}r{1pt}){6-8}
    Sev & $S_{\bm{1}}$ & $S_{\bm{2}}$ & $S_{\bm{3}}$ & & U$_{\bm{1}}$ & U$_{\bm{2}}$ & U$_{\bm{3}}$ \\
    \midrule
    1    & \ldotsbox & \emptybox & \hatchbox &  & \ldotsbox & \emptybox & \emptybox \\
    2    & \ldotsbox & \hatchbox & \hatchbox &  & \ldotsbox & \hatchbox & \emptybox \\
    3    & \ldotsbox & \hatchbox & \hatchbox &  & \ldotsbox & \ldotsbox & \emptybox \\
    5    & \emptybox & \hatchbox & \hatchbox &  & \emptybox & \hatchbox & \hatchbox \\
    10    & \hatchbox & \hatchbox & \hatchbox &  & \emptybox & \hatchbox & \hatchbox \\
    20    & \hatchbox & \hatchbox & \hatchbox &  & \emptybox & \ldotsbox & \hatchbox \\
    50   & \hatchbox & \hatchbox & \hatchbox &  & \hatchbox & \hatchbox & \hatchbox \\
    100   & \hatchbox & \hatchbox & \hatchbox &  & \emptybox & \hatchbox & \hatchbox \\
    \bottomrule
    \end{tabularx}

    \vspace{2pt}

    \begin{tablenotes}
      \setlength{\itemsep}{0.3em} %
      \item \raisebox{-0.5ex}{\emptybox}\hspace{4pt} no performance change
      \item \raisebox{-0.5ex}{\hatchbox}\hspace{4pt} performance regression
      \item \raisebox{-0.5ex}{\ldotsbox}\hspace{4pt} performance increase
    \end{tablenotes}

    \end{threeparttable}
    \label{tab:results}
\end{table}

\subsubsection*{Instrumentation Overhead}
\label{sec:overhead}
To quantify instrumentation overhead, we conduct A/A experiments in which we instrument one version but not the other.
We repeat the experiment 5 times on different machines and collect roughly $51.6$k request-latency measurements overall across the three endpoints.

\begin{figure*}[!t]
    \centering
    \includegraphics[width=\textwidth]{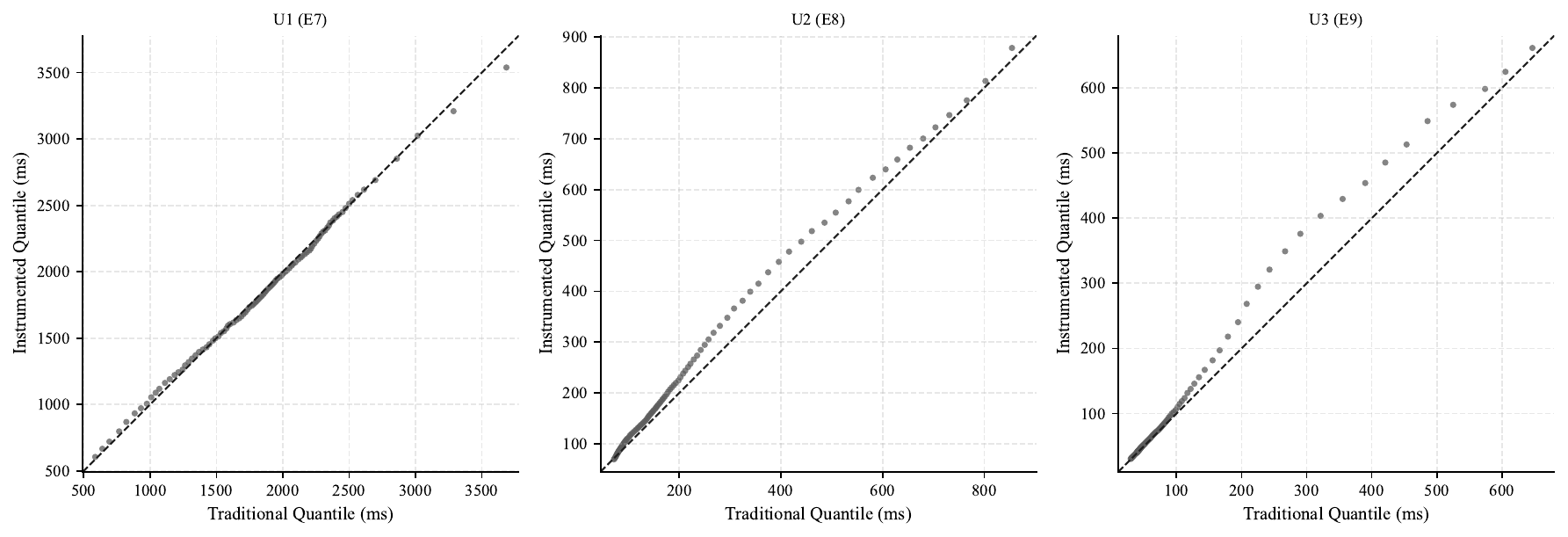}
    \caption{Q-Q Plot for A/A experiments per endpoint. The closer the points are to the diagonal, the more similar the distributions are.}
    \label{fig:overhead}
  \end{figure*}

\cref{fig:overhead} shows the Q-Q plots for all endpoints with one percentile as the step size.
The distributions of both the traditional and instrumented approach are fairly similar with differences mostly in the higher percentiles. 
We attribute these to infrastructure effects and the high load of the system during our experiments.
Due to the nature of the application, versatile workloads, and random fluctuations, we do observe performance changes at the external interfaces.
This, however, highlights the importance of the internal instrumentation.

The Hellinger distance is a good indicator of the similarity of the two distributions.
Values below 0.1 indicate a very similar distribution, values between 0.1 and 0.2 indicate a moderate similarity and values above 0.2 indicate a significant difference.
For the search operation, we compute a distance of 0.080, for the ranking operation 0.063 and for the recommendation operation 0.048.

\section{Discussion}
\label{sec:discussion}
We introduce \sysname as a novel approach to sensitive performance change detection in application benchmarks.
However, our approach has limitations and challenges that we need to address in future work.

\subsubsection*{No Production Setup \& Limited Scope}
We inject artificial performance regressions, particularly to evaluate sensitivity with different severity levels.
We design our testbed application to be realistic; however, it is not used in production, and correspondingly, our experiments have a limited scope.

Nevertheless, we expect \sysname to be even more relevant for increasingly complex applications, where performance characteristics become harder to understand at external interfaces.

\subsubsection*{Performance Change Coverage}
We face the challenge that code changes are versatile, and our prototype does not handle all change types well.
While we can show high localization quality, particularly for recall, in our experiments, the question remains how well we can generalize this to real-world applications.
Especially complex code changes that are hard to understand, even for humans, can be missed as the reasoning capabilities are not sufficient.
Further, performance behavior might be hidden or externally influenced by, e.g., environment variables, which we consider an anti-pattern and is out of scope for our approach.
Nevertheless, it is important that \sysname does not significantly alter performance characteristics on the external APIs of a service.
Hidden or missed performance changes should still be visible at the external interfaces.
In our experiments, we did not observe a strong influence on the performance behavior at the external interfaces.

\subsubsection*{Scalability}
We recognize scalability in terms of cost and time as a major challenge and possible limitation of our approach.

For every benchmarking experiment, the instrumentation procedure runs first, which consumes time and money.
In our experiments, we limit the number of changes between service versions to a small fraction of what we expect in real-world applications, particularly in complex microservice architectures.
Running our full procedure for potentially thousands of code changes might introduce high costs and instrumentation times.
Therefore, we handle code changes in parallel in multiple threads, since processing the effects of a single change might be time-consuming.
Handling files in parallel keeps the overall duration low.
We can extend our system design to support even higher parallelization through distributed compute environments, e.g., Function-as-a-Service (FaaS) instead of threads, but keep that for future work.
To enhance cost-efficiency, we collect data during the instrumentation step, which we can use to train smaller and cheaper models, a process commonly referred to as model distillation.
We leave using purpose-tuned open-weight models for future work.

\subsubsection*{Language Support}
We can expand our approach to different programming languages, but for the scope of this paper, we limit ourselves to JavaScript and TypeScript.
Node.js provides the AsyncLocalStorage API for request context sharing, making it easy to add nested spans without complex code changes.
Python provides similar concepts, such as contextvars. 
Other programming languages without convenient request context sharing might be harder to support, but generally, it is feasible to generalize language support.

\subsubsection*{Finding the Right Level of Sensitivity}
We support a sensitivity parameter that determines how sensitively we react to code changes.
With lower sensitivity, we skip more code changes, which helps if we are only interested in major performance changes.
We provide a reasonable default level that balances precision and recall.
Finding the right detail level, particularly optimized for a given use case, remains a challenge and is out of scope for this paper.

\section{Conclusion \& Future Work}
Detecting performance changes in application benchmarks is challenging because traditional end-to-end measurements are often too insensitive to detect small regressions that cause performance drifts over time.

We presented \sysname, which integrates change-localized lightweight performance measurements into application benchmarks.
Through these change-localized measurements, combined with duet benchmarking, our approach substantially improves performance change detection compared to traditional benchmarks, detecting regressions at up to 5$\times$ lower injected severity while preserving similar A/A latency distributions (Hellinger distance 0.048--0.080).
In our experiments, instrumentation quality is high (58\% precision, 93\% recall, and 71\% specificity, averaged across tasks, with a line-distance threshold of five), and we do not observe significant changes in performance behavior at external interfaces, i.e., our approach appears to have no prominent downsides when the instrumentation itself is reliable, fast, and scalable.

Our prototype implementation is a proof of concept that demonstrates the feasibility of \sysname.
\Sysname is agnostic to the application design and can be adapted with a few constraints.

As future work, it will be particularly interesting to apply it in more complex microservice architectures with distributed tracing using open-weight models.
Since we closely follow the OpenTelemetry methodology, we expect \sysname to integrate easily into existing systems and microservices.

\balance
\bibliographystyle{IEEEtran}
\bibliography{bibliography}

\begin{thebibliography}{10}
\providecommand{\url}[1]{#1}
\csname url@samestyle\endcsname
\providecommand{\newblock}{\relax}
\providecommand{\bibinfo}[2]{#2}
\providecommand{\BIBentrySTDinterwordspacing}{\spaceskip=0pt\relax}
\providecommand{\BIBentryALTinterwordstretchfactor}{4}
\providecommand{\BIBentryALTinterwordspacing}{\spaceskip=\fontdimen2\font plus
\BIBentryALTinterwordstretchfactor\fontdimen3\font minus
  \fontdimen4\font\relax}
\providecommand{\BIBforeignlanguage}[2]{{%
\expandafter\ifx\csname l@#1\endcsname\relax
\typeout{** WARNING: IEEEtran.bst: No hyphenation pattern has been}%
\typeout{** loaded for the language `#1'. Using the pattern for}%
\typeout{** the default language instead.}%
\else
\language=\csname l@#1\endcsname
\fi
#2}}
\providecommand{\BIBdecl}{\relax}
\BIBdecl

\bibitem{schermann2016bifrost}
G.~Schermann, D.~Sch{\"o}ni, P.~Leitner, and H.~C. Gall, ``Bifrost: Supporting
  continuous deployment with automated enactment of multi-phase live testing
  strategies,'' in \emph{Proceedings of the 17th International Middleware
  Conference}, 2016.

\bibitem{continuous_benchmarking_grambow}
M.~Grambow, F.~Lehmann, and D.~Bermbach, ``Continuous benchmarking: Using
  system benchmarking in build pipelines,'' in \emph{2019 IEEE International
  Conference on Cloud Engineering (IC2E)}, 2019, pp. 241--246.

\bibitem{continuous_benchmarking_daly_2021}
\BIBentryALTinterwordspacing
D.~Daly, ``Creating a virtuous cycle in performance testing at mongodb,'' in
  \emph{Proceedings of the ACM/SPEC International Conference on Performance
  Engineering}, ser. ICPE '21.\hskip 1em plus 0.5em minus 0.4em\relax ACM, Apr.
  2021. [Online]. Available: \url{http://dx.doi.org/10.1145/3427921.3450234}
\BIBentrySTDinterwordspacing

\bibitem{leitner2016patternschaosstudy}
\BIBentryALTinterwordspacing
P.~Leitner and J.~Cito, ``Patterns in the chaos - {A} study of performance
  variation and predictability in public iaas clouds,'' \emph{{ACM}
  Transactions on Internet Technology}, vol.~16, no.~3, pp. 15:1--15:23, 2016.
  [Online]. Available: \url{https://doi.org/10.1145/2885497}
\BIBentrySTDinterwordspacing

\bibitem{Bermbach_book_2017}
D.~Bermbach, E.~Wittern, and S.~Tai, \emph{Cloud Service Benchmarking:
  Measuring Quality of Cloud Services from a Client Perspective}, 1st~ed.\hskip
  1em plus 0.5em minus 0.4em\relax Springer Publishing Company, Incorporated,
  2017.

\bibitem{Japke_2023}
\BIBentryALTinterwordspacing
N.~Japke, C.~Witzko, M.~Grambow, and D.~Bermbach, ``The early microbenchmark
  catches the bug -- studying performance issues using micro- and application
  benchmarks,'' in \emph{Proceedings of the 16th IEEE/ACM International
  Conference on Utility and Cloud Computing}, ser. UCC '23.\hskip 1em plus
  0.5em minus 0.4em\relax New York, NY, USA: Association for Computing
  Machinery (ACM), Dec. 2023. [Online]. Available:
  \url{https://doi.org/10.1145/3603166.3632128}
\BIBentrySTDinterwordspacing

\bibitem{Grambow_2023}
\BIBentryALTinterwordspacing
M.~Grambow, D.~Kovalev, C.~Laaber, P.~Leitner, and D.~Bermbach, ``Using
  microbenchmark suites to detect application performance changes,'' \emph{IEEE
  Transactions on Cloud Computing}, vol.~11, no.~3, pp. 2575--2590, Jul. 2023.
  [Online]. Available: \url{http://dx.doi.org/10.1109/TCC.2022.3217947}
\BIBentrySTDinterwordspacing

\bibitem{missing_performance_tests}
\BIBentryALTinterwordspacing
P.~Stefan, V.~Horky, L.~Bulej, and P.~Tuma, ``Unit testing performance in java
  projects: Are we there yet?'' in \emph{Proceedings of the 8th ACM/SPEC on
  International Conference on Performance Engineering}, ser. ICPE '17.\hskip
  1em plus 0.5em minus 0.4em\relax New York, NY, USA: Association for Computing
  Machinery, 2017, pp. 401--412. [Online]. Available:
  \url{https://doi.org/10.1145/3030207.3030226}
\BIBentrySTDinterwordspacing

\bibitem{Schirmer_2023}
\BIBentryALTinterwordspacing
T.~Schirmer, N.~Japke, S.~Greten, T.~Pfandzelter, and D.~Bermbach, ``The night
  shift: Understanding performance variability of cloud serverless platforms,''
  in \emph{Proceedings of the 1st Workshop on SErverless Systems, Applications
  and MEthodologies}, ser. SESAME ’23.\hskip 1em plus 0.5em minus 0.4em\relax
  ACM, May 2023, p. 27–33. [Online]. Available:
  \url{http://dx.doi.org/10.1145/3592533.3592808}
\BIBentrySTDinterwordspacing

\bibitem{Laaber_2019}
\BIBentryALTinterwordspacing
C.~Laaber, J.~Scheuner, and P.~Leitner, ``Software microbenchmarking in the
  cloud. how bad is it really?'' \emph{Empirical Software Engineering},
  vol.~24, no.~4, pp. 2469--2508, Aug. 2019. [Online]. Available:
  \url{https://doi.org/10.1007/s10664-019-09681-1}
\BIBentrySTDinterwordspacing

\bibitem{Bulej_2020_Duet_Benchmarking}
\BIBentryALTinterwordspacing
L.~Bulej, V.~Horký, P.~Tuma, F.~Farquet, and A.~Prokopec, ``Duet benchmarking:
  Improving measurement accuracy in the cloud,'' in \emph{Proceedings of the
  ACM/SPEC International Conference on Performance Engineering}, ser. ICPE
  ’20.\hskip 1em plus 0.5em minus 0.4em\relax ACM, Apr. 2020, pp. 100--107.
  [Online]. Available: \url{http://dx.doi.org/10.1145/3358960.3379132}
\BIBentrySTDinterwordspacing

\bibitem{RMIT}
\BIBentryALTinterwordspacing
A.~Abedi and T.~Brecht, ``Conducting repeatable experiments in highly variable
  cloud computing environments,'' in \emph{Proceedings of the 8th ACM/SPEC on
  International Conference on Performance Engineering}, ser. ICPE '17.\hskip
  1em plus 0.5em minus 0.4em\relax New York, NY, USA: Association for Computing
  Machinery, 2017, p. 287–292. [Online]. Available:
  \url{https://doi.org/10.1145/3030207.3030229}
\BIBentrySTDinterwordspacing

\bibitem{Davison_Hinkley_1997}
A.~C. Davison and D.~V. Hinkley, \emph{Bootstrap Methods and their
  Application}, ser. Cambridge Series in Statistical and Probabilistic
  Mathematics.\hskip 1em plus 0.5em minus 0.4em\relax Cambridge University
  Press, 1997.

\bibitem{schirmer2024elastibench}
T.~Schirmer, T.~Pfandzelter, and D.~Bermbach, ``Elastibench: Scalable
  continuous benchmarking on cloud faas platforms,'' in \emph{Proceedings of
  the 12th IEEE International Conference on Cloud Engineering}, ser. IC2E
  '24.\hskip 1em plus 0.5em minus 0.4em\relax New York, NY, USA: IEEE, Sep.
  2024.

\bibitem{costa2019s}
D.~Costa, C.-P. Bezemer, P.~Leitner, and A.~Andrzejak, ``What's wrong with my
  benchmark results? studying bad practices in jmh benchmarks,'' \emph{IEEE
  Transactions on Software Engineering}, vol.~47, no.~7, pp. 1452--1467, 2019.

\bibitem{japke2025muoptimestaticallyreducingexecution}
\BIBentryALTinterwordspacing
N.~Japke, M.~Grambow, C.~Laaber, and D.~Bermbach, ``{$\mu$OpTime}: Statically
  reducing the execution time of microbenchmark suites using stability
  metrics,'' \emph{{ACM} Transactions on Software Engineering and Methodology},
  Jan. 2025. [Online]. Available: \url{https://doi.org/10.1145/3715322}
\BIBentrySTDinterwordspacing

\bibitem{bulej2019initial}
\BIBentryALTinterwordspacing
L.~Bulej, V.~Hork{\'{y}}, and P.~Tuma, ``Initial experiments with duet
  benchmarking: Performance testing interference in the cloud,'' in \emph{27th
  {IEEE} International Symposium on Modeling, Analysis, and Simulation of
  Computer and Telecommunication Systems, {MASCOTS} 2019, Rennes, France,
  October 21-25, 2019}.\hskip 1em plus 0.5em minus 0.4em\relax {IEEE} Computer
  Society, 2019, pp. 249--255. [Online]. Available:
  \url{https://doi.org/10.1109/MASCOTS.2019.00035}
\BIBentrySTDinterwordspacing

\bibitem{japke2025isolation}
\BIBentryALTinterwordspacing
N.~Japke, F.~Hamdan, D.~Baumann, and D.~Bermbach, ``Investigating the impact of
  isolation on synchronized benchmarks,'' in \emph{Proceedings of the 18th
  IEEE/ACM International Conference on Utility and Cloud Computing}, ser. UCC
  '25.\hskip 1em plus 0.5em minus 0.4em\relax New York, NY, USA: Association
  for Computing Machinery, 2026. [Online]. Available:
  \url{https://doi.org/10.1145/3773274.3774703}
\BIBentrySTDinterwordspacing

\bibitem{Reichelt_2022_Automated_Code_Level_Performance_Changes}
\BIBentryALTinterwordspacing
D.~G. Reichelt, S.~Kühne, and W.~Hasselbring, ``Automated identification of
  performance changes at code level,'' in \emph{2022 IEEE 22nd International
  Conference on Software Quality, Reliability and Security (QRS)}.\hskip 1em
  plus 0.5em minus 0.4em\relax IEEE, Dec. 2022, pp. 916--925. [Online].
  Available: \url{http://dx.doi.org/10.1109/QRS57517.2022.00096}
\BIBentrySTDinterwordspacing

\bibitem{Chen_JIT_Performance_Regression}
\BIBentryALTinterwordspacing
J.~Chen, W.~Shang, and E.~Shihab, ``{ PerfJIT: Test-Level Just-in-Time
  Prediction for Performance Regression Introducing Commits },'' \emph{IEEE
  Transactions on Software Engineering}, vol.~48, no.~05, pp. 1529--1544, May
  2022. [Online]. Available:
  \url{https://doi.ieeecomputersociety.org/10.1109/TSE.2020.3023955}
\BIBentrySTDinterwordspacing

\bibitem{waller_including_2015}
\BIBentryALTinterwordspacing
J.~Waller, N.~C. Ehmke, and W.~Hasselbring, ``Including performance benchmarks
  into continuous integration to enable devops,'' \emph{ACM SIGSOFT Software
  Engineering Notes}, vol.~40, no.~2, pp. 1--4, apr 2015. [Online]. Available:
  \url{https://doi.org/10.1145/2735399.2735416}
\BIBentrySTDinterwordspacing

\bibitem{javed_perfci_2020}
\BIBentryALTinterwordspacing
O.~Javed, J.~H. Dawes, M.~Han, G.~Franzoni, A.~Pfeiffer, G.~Reger, and
  W.~Binder, ``Perfci: A toolchain for automated performance testing during
  continuous integration of python projects,'' in \emph{Proceedings of the 35th
  IEEE/ACM International Conference on Automated Software Engineering}, ser.
  ASE '20.\hskip 1em plus 0.5em minus 0.4em\relax New York, NY, USA:
  Association for Computing Machinery, 2021, pp. 1344--1348. [Online].
  Available: \url{https://doi.org/10.1145/3324884.3415288}
\BIBentrySTDinterwordspacing

\bibitem{daly_industry_2019}
\BIBentryALTinterwordspacing
D.~Daly, W.~Brown, H.~Ingo, J.~O'Leary, and D.~Bradford, ``The use of change
  point detection to identify software performance regressions in a continuous
  integration system,'' in \emph{Proceedings of the ACM/SPEC International
  Conference on Performance Engineering}, ser. ICPE '20.\hskip 1em plus 0.5em
  minus 0.4em\relax New York, NY, USA: Association for Computing Machinery,
  2020, pp. 67--75. [Online]. Available:
  \url{https://doi.org/10.1145/3358960.3375791}
\BIBentrySTDinterwordspacing

\bibitem{LaaberMBEval}
\BIBentryALTinterwordspacing
C.~Laaber and P.~Leitner, ``An evaluation of open-source software
  microbenchmark suites for continuous performance assessment,'' in
  \emph{Proceedings of the 15th International Conference on Mining Software
  Repositories}, ser. MSR '18.\hskip 1em plus 0.5em minus 0.4em\relax New York,
  NY, USA: Association for Computing Machinery, May 2018, pp. 119--130.
  [Online]. Available: \url{https://doi.org/10.1145/3196398.3196407}
\BIBentrySTDinterwordspacing

\bibitem{japke2025cicd}
N.~Japke, S.~Koch, H.~Lukasczyk, and D.~Bermbach, ``Towards an optimized
  benchmarking platform for ci/cd pipelines,'' in \emph{2025 IEEE International
  Conference on Cloud Engineering (IC2E)}, 2025, pp. 36--41.

\bibitem{grambow2021befaasapplicationcentricbenchmarkingframework}
M.~Grambow, T.~Pfandzelter, L.~Burchard, C.~Schubert, M.~Zhao, and D.~Bermbach,
  ``{BeFaaS}: An application-centric benchmarking framework for faas
  platforms,'' in \emph{2021 IEEE International Conference on Cloud Engineering
  (IC2E)}, 2021, pp. 1--8.

\bibitem{Yao_2023_ReAct}
\BIBentryALTinterwordspacing
S.~Yao, J.~Zhao, D.~Yu, N.~Du, I.~Shafran, K.~Narasimhan, and Y.~Cao, ``React:
  Synergizing reasoning and acting in language models,'' 2023. [Online].
  Available: \url{https://arxiv.org/abs/2210.03629}
\BIBentrySTDinterwordspacing

\bibitem{instrumentation_overhead}
\BIBentryALTinterwordspacing
D.~G. Reichelt, L.~Bulej, R.~Jung, and A.~van Hoorn, ``Overhead comparison of
  instrumentation frameworks,'' in \emph{Companion of the 15th ACM/SPEC
  International Conference on Performance Engineering}, ser. ICPE '24
  Companion.\hskip 1em plus 0.5em minus 0.4em\relax New York, NY, USA:
  Association for Computing Machinery, 2024, pp. 249--256. [Online]. Available:
  \url{https://doi.org/10.1145/3629527.3652269}
\BIBentrySTDinterwordspacing

\bibitem{octoverse2025}
G.~Staff, ``Octoverse: A new developer joins github every second as ai leads
  typescript to \#1,''
  \url{https://github.blog/news-insights/octoverse/octoverse-a-new-developer-joins-github-every-second-as-ai-leads-typescript-to-1/},
  2025, updated November 7, 2025. Accessed January 27, 2026.

\bibitem{hastie2014matrixcompletionlowranksvd}
\BIBentryALTinterwordspacing
T.~Hastie, R.~Mazumder, J.~Lee, and R.~Zadeh, ``Matrix completion and low-rank
  svd via fast alternating least squares,'' 2014. [Online]. Available:
  \url{https://arxiv.org/abs/1410.2596}
\BIBentrySTDinterwordspacing

\bibitem{pagerank}
\BIBentryALTinterwordspacing
L.~Page, S.~Brin, R.~Motwani, and T.~Winograd, ``The pagerank citation ranking:
  Bringing order to the web.'' Stanford InfoLab, Technical Report 1999-66,
  November 1999, previous number = SIDL-WP-1999-0120. [Online]. Available:
  \url{http://ilpubs.stanford.edu:8090/422/}
\BIBentrySTDinterwordspacing

\end{thebibliography}

\end{document}